\documentclass[12pt]{iopart}

\usepackage{latexsym, graphicx}
\usepackage{color, hyperref}

\begin{document}

\title{Quantum teleportation and entanglement swapping with long baseline in outer space}

\author{Shih-Yuin Lin$^1$ and B. L. Hu$^2$}
\address{$^1$ Department of Physics, National Changhua University of Education, Changhua 50007, Taiwan}
\address{$^2$ Joint Quantum Institute and Maryland Center for Fundamental Physics, \\
University of Maryland, College Park, Maryland 20742-4111, USA}
\ead{sylin@cc.ncue.edu.tw and blhu@umd.edu}
\begin{indented}
\item[]December 8, 2020
\end{indented}

\begin{abstract}
Quantum information experiments applying quantum optics in outer space with a very long baseline may have advantages over the current earth-bound experiments or the earth-to-satellite experiments because they can minimize the loss in light transmission and maximize the gain in time resolution. This future class of experiments, amongst them quantum teleportation and entanglement swapping, can shed light on many fundamental theoretical issues in gravitational quantum physics and relativistic quantum information. Regarding relativity theory, these experiments in an outer-space setting can involve observers at spacelike and timelike separations and explicate intriguing phenomena from different choices of time-slicing. Regarding quantum information, they may be able to ensure the causal independence of the expectation values in the Bell test. These issues are addressed in this paper with analysis and explanations.
\end{abstract}

\maketitle
 
\section{Introduction}
\label{DSQLadv}

Long-distance quantum information (QI) transmission is the key element in establishing a global quantum network. Much effort has been made in the ground-to-satellite QI experiments transmitting photons with quantum coherence \cite{RT02, UJ09, GK17, LC18}. 
Recently, entanglement distribution by a moving satellite to distant ground bases (where the Bell test is performed) \cite{YP17} and quantum teleportation from a ground base to the satellite have been implemented \cite{RP17}. In the near future, further development of technology may enable similar experiments at the planetary scale in outer space. Not only would this be a natural extension of the current ground-satellite QI technology to a much larger scale, but it ushers in a new regime of testing important issues in quantum foundations.

\begin{figure}
\includegraphics[width=4.4cm]{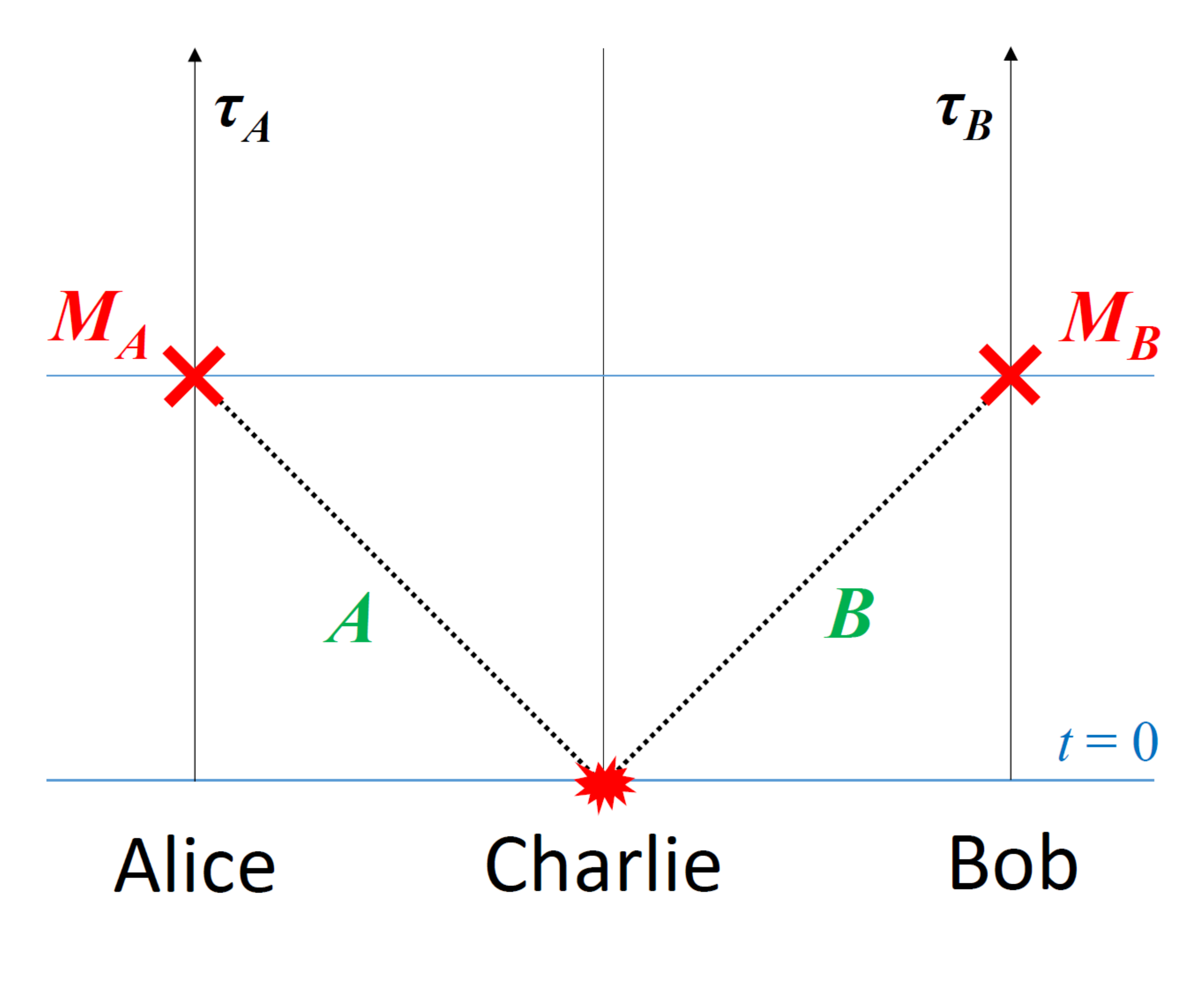}\hspace{.2cm}
\includegraphics[width=5.4cm]{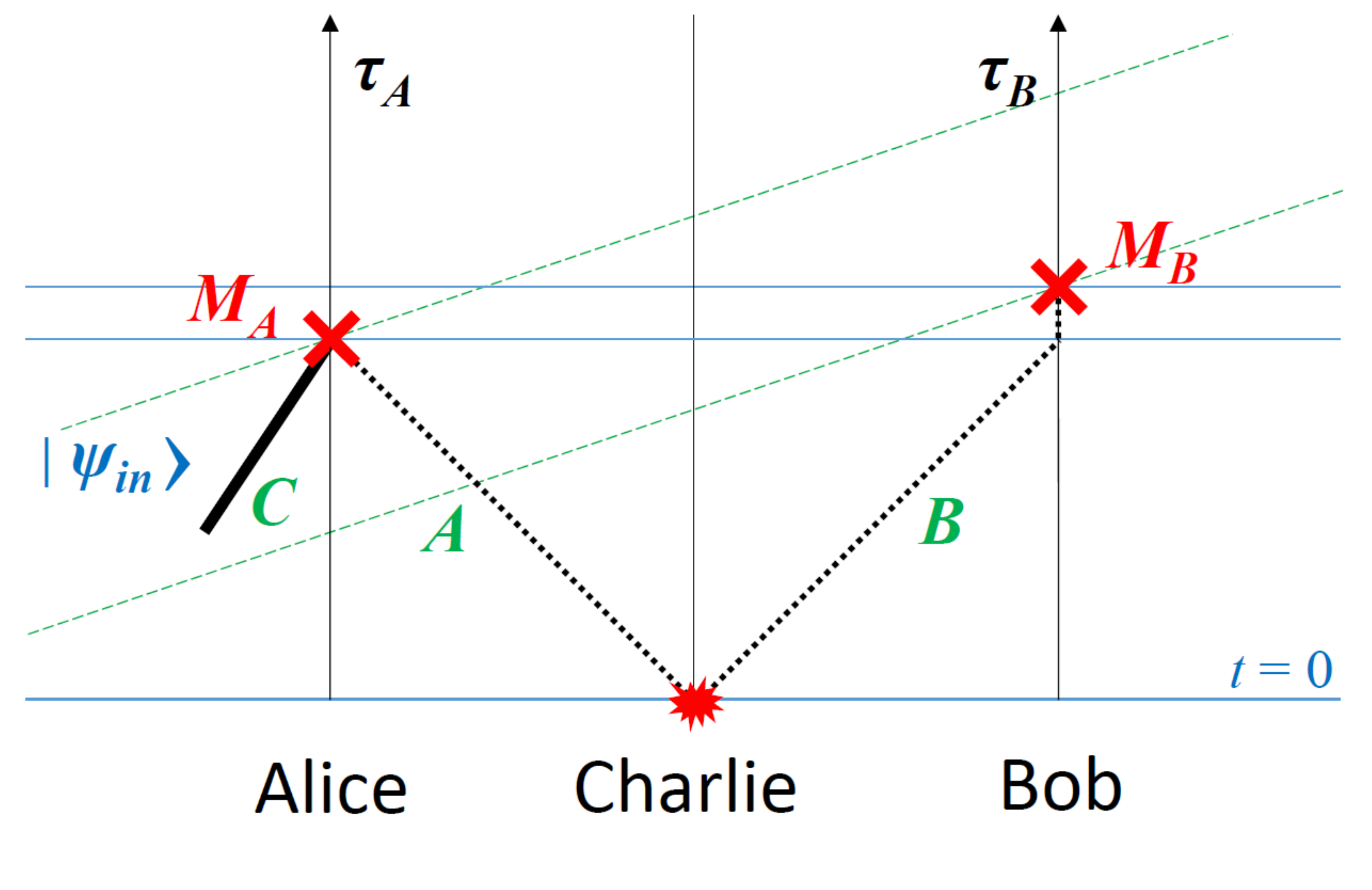}\hspace{.2cm}
\includegraphics[width=6cm]{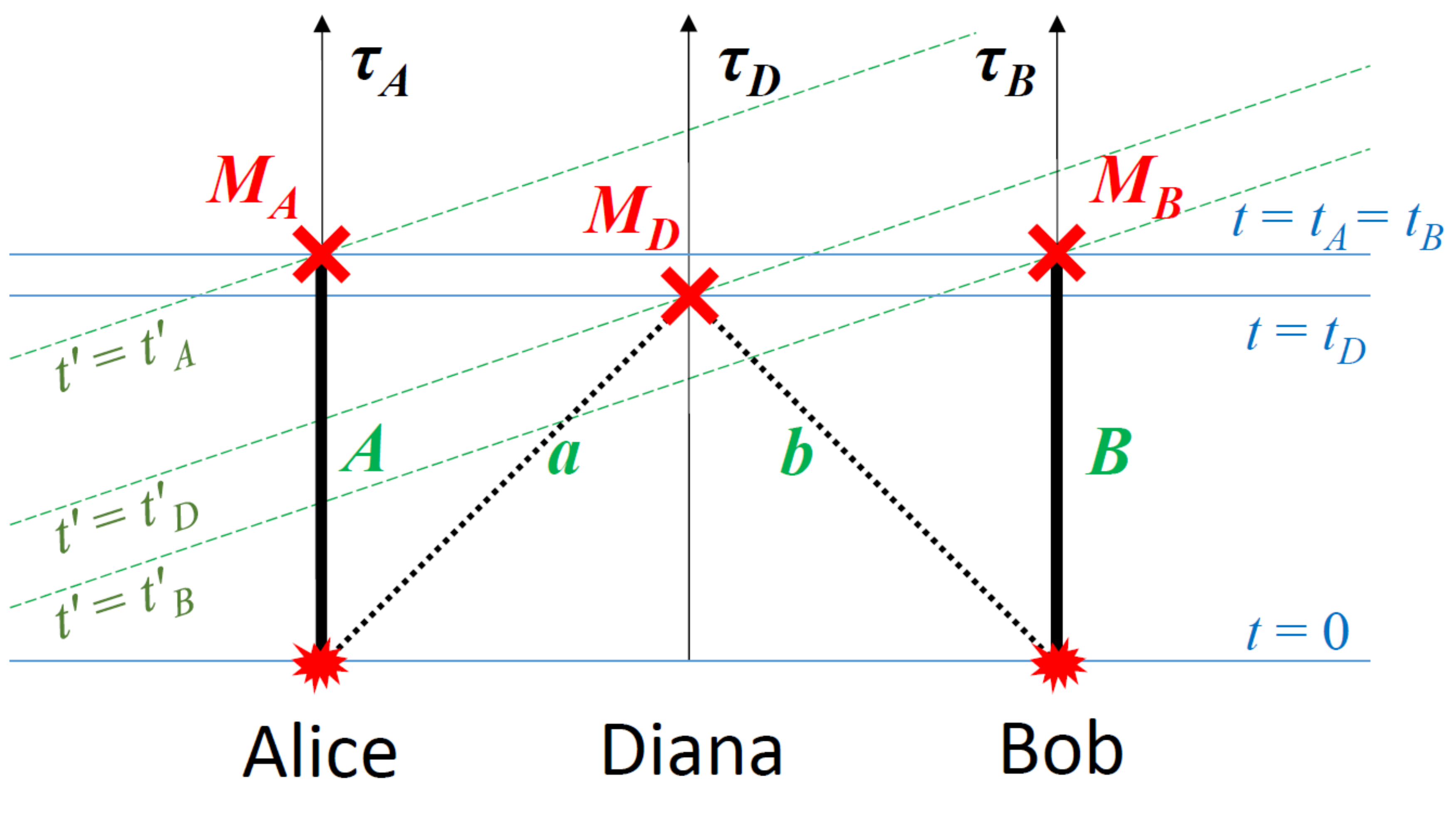}
\caption{Spacetime diagrams of the Bell test (left), the incomplete ``passive" quantum teleportation (middle) [for complete quantum teleportation, see Figure \ref{EntSwapQTel})] and entanglement swapping (right), which involve two ($A$ and $B$), three ($A$, $B$, and $C$), and four ($A$, $B$, $a$, and $b$) quantum-state carriers, respectively. Here, the red ``$\times$" represents local measurement events,
the black dotted lines and thick solid lines represent the worldlines of the participating quantum objects, the blue solid lines (green dotted lines) represent the time slices in the coordinates of some bookkeeper (some alternative bookkeeper moving in a different way), and the blue dashed line in the middle plot represents the light signal carrying the classical information $I_A$ of the measurement outcome by Alice.}
\label{NonLocalExpt}
\end{figure}

In this paper we consider QI experiments of a very long baseline in space, specifically, quantum teleportation and entanglement swapping (Figures \ref{NonLocalExpt} and \ref{EntSwapQTel}) between Alice on the International Space Station (ISS) and Bob on the Luner Gateway (LG) \cite{ISS20, LG20, DSQL1}. The ISS is orbiting the Earth at an altitude about 400 km above the sea level at a centripetal acceleration $8.7 {\rm m/s^2}$, with the period about 1/15.5 days $\approx 5.6 \times 10^3$ s \cite{ISS20}. The LG would be orbiting the Moon along the Near Rectilinear Halo Orbit (NRHO) with the periapsis about 3000 km and the apoapsis about 70000 km to the Moon, and the period about 7 days $\approx 6 \times 10^5$ s \cite{LG20}. 
An obvious advantage of such kind of experiments in space is that photons traveling between observers in outer space could survive better from scattering than those traversing the Earth's atmosphere at the same distance. This means that quantum optics-based QI  experiments with a baseline much longer than what we have on the surface of the Earth may be feasible. 

In the experiments revealing the quantum advantages of a nonlocal system, the difference from the ``classical" results with local hidden variables can not be seen in a single-shot measurement but only at the level of probabilities or expectation values, such as the Bell-like inequalities in the Bell test \cite{Bell, CHSH, GS15} (e.g., (\ref{CHSHineq}) and (\ref{CHSHBell})), the averaged fidelity of quantum teleportation \cite{BB93, PE15} (e.g., (\ref{FiQT}), (\ref{rin}), and (\ref{FiQTav})), and the expectation values for the entanglement witness \cite{HZ15} in entanglement swapping \cite{Pe00, JZ02, MZ12} (Figure \ref{NonLocalExpt}).
To obtain those probabilities or expectation values, a series of the same processes must be repeated many times in the same experiment so the bookkeeper can collect enough outcomes for performing the statistics
\footnote{Our ``bookkeeper" is defined in the sense as in Ref. \cite{TW00}, referring to an archivist available for consultation at all times, who does not make physical observation but can collect the outcomes of all the measurements from local physical observers (in reality, there are no nonlocal observers) on a system, coordinatize all the events of measurement and operation on that system, 
calculate the probability, expectation values, and correlators of the physical observables, and then describe the evolution of the observed system accordingly. One local observer $A$ cannot tell directly what another local observer $B$ sees or measures, but can use the records kept by the bookkeeper from the relevant observers involved to work this out by some mathematics and then draw the inference.  Confusion in the description of nonlocal events (such as the ``before-before" kind -- see 
\ref{BBAA}) is often generated when one inadvertently switches over to the ``global" view of this bookkeeper not realizing that it is not from a local physical observer.}.

\begin{figure}
  \includegraphics[width=4cm]{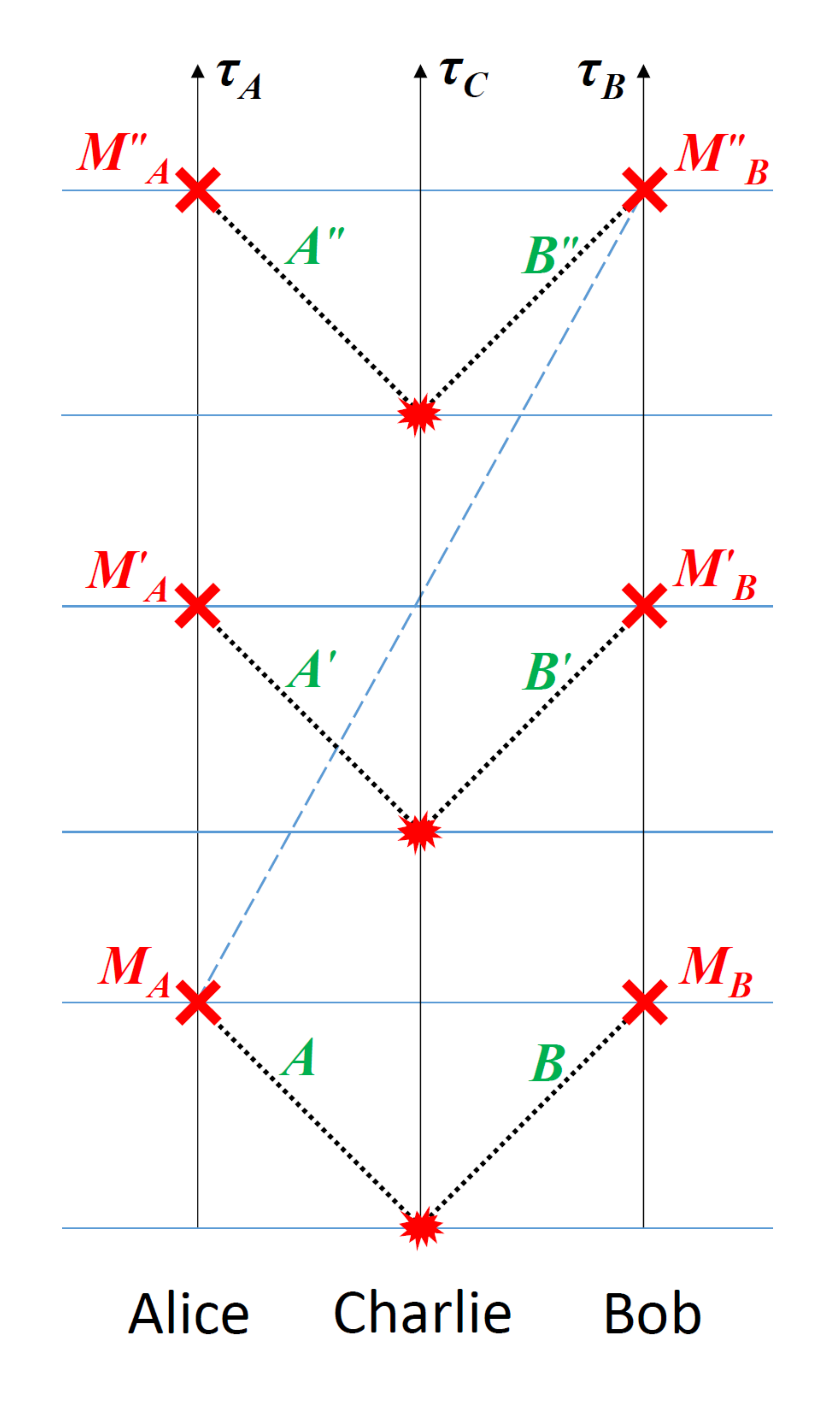} 
  \includegraphics[width=5cm]{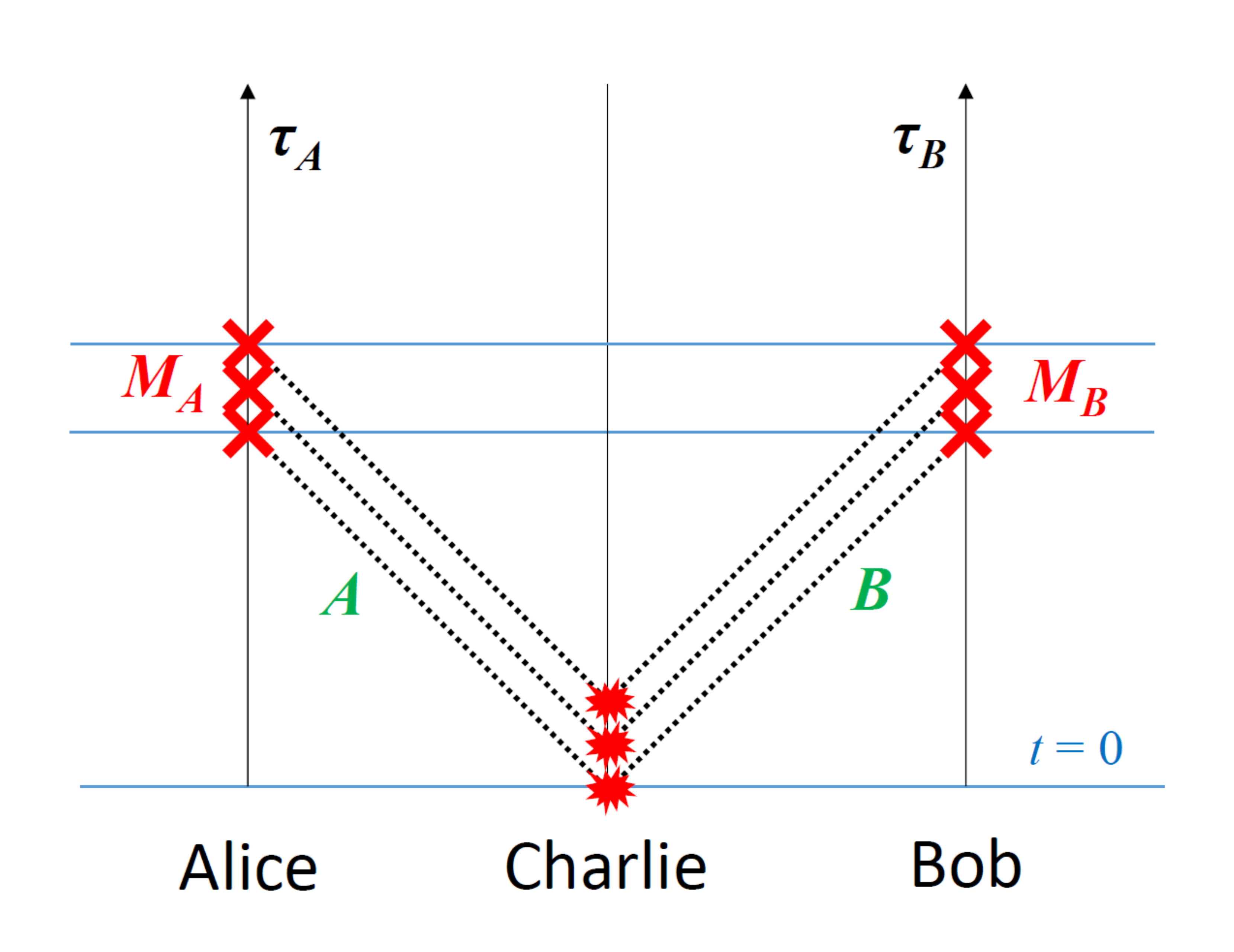}  
	\includegraphics[width=5.5cm]{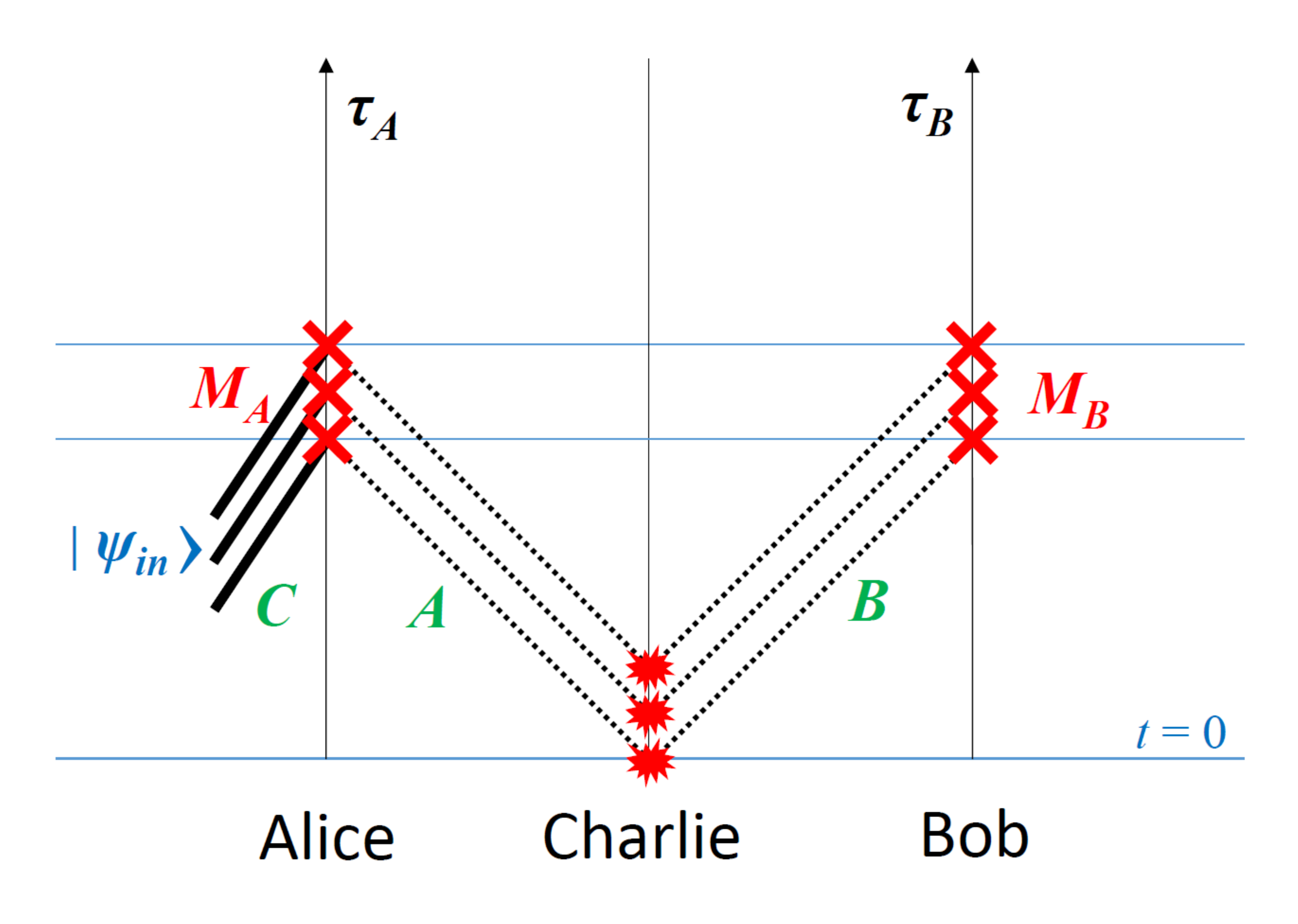}  
\caption{In a conventional experiment of the Bell test (left), a series of identical processes are carried out successively while a later measurement event can be in the future lightcone of earlier events (e.g., $M''_B$ and $M_A$). Between the ISS and LG one may be able to achieve the Bell test (middle) 
and ``passive" quantum teleportation (right) 
with all of Alice's measurements $M_A$ spacelike separated from all of Bob's measurements $M_B$ in the period of sampling.} 
\label{BellQTel}
\end{figure}

In conventional ground-based experiments, because of the limited separation, a later event in one agent's worldline would be in the future lightcones of some earlier events in both agents' worldlines (e.g., \cite{GS15}), and so those events are causally connected [e.g., the events $M_B''$ and $M_A$ in Figure \ref{BellQTel} (left)]. In the Bell test, this corresponds to the memory loophole \cite{Barrett02, Gill01, Larsson14} which can be suppressed if the number of the measurement outcomes is sufficiently large in statistics \footnote{Statistical fluctuations may violate the CHSH inequalities in a quantum information experiment of finite runs. In Ref.\cite{Barrett02}, the violation of an alternative form of the CHSH inequality is shown to be bounded by $5N^{-1/2 + \epsilon}+ 5 \sqrt{3/2 \pi} N^{-\epsilon} \exp (-N^{2\epsilon}/6)$, where $N$ is the total number of the successful runs and $\epsilon$ is a small positive number. For a sufficiently large $N$, this violation will be suppressed. However, for a very small $\epsilon$, the second term in the bound ($\sim N^{-\epsilon} \exp (-N^{2\epsilon}/6)$) will dominate over the first term ($\sim N^{-1/2 + \epsilon}$) in a large interval of $N$ where $N$ is not too small or too large, and one has to take a very large $N$ to see the suppression of the bound.}.
To examine if one can really suppress the memory loophole, particularly the one-sided and simultaneous memory loopholes where the collection of all measurements by one agent (Alice) is spacelike separated from the collection by the other agent (Bob) \cite{Barrett02}, a very long baseline for the Bell test and the related QI experiments is needed. 
We expect that the separation of the ISS and the LG could accommodate the trains of photons long enough to have all the measurements around $M_A$ spacelike-separated from those around $M_B$ [Figure \ref{BellQTel} (middle and right) and Figure \ref{EntSwapDLater}] and so we may be able to test with a cleaner causal independence of the outcomes of different groups for statistics. 
At this length scale, one may also check if a similar loophole is present in quantum teleportation or entanglement swapping [Figure \ref{BellQTel} (middle) and (right)].

When multiplied by such a large length scale, however, even a tiny beam divergence 
can cause a serious loss of photons, provided that the sizes of the receiver and emitter apertures are of the same order. The coincidence counts of photons at different sites in the proposed experiments could be so rare that statistical fluctuations and the classical correlations potentially introduced in data selection would 
dominate over the quantum correlations we are looking for. 
In the Bell test, the detector-efficiency loophole and the coincidence-time loophole \cite{La98, LG04} may become significant due to the high loss of photons with this long baseline. While these loopholes could be fixed by considering the Clauser-Horne (CH) inequality \cite{CH74} instead of the CHSH inequality (\ref{CHSHineq}) \cite{LG14} in the Bell test, 
similar loopholes may also make a classical result appears quantum 
in quantum teleportation and entanglement swapping.

Below, we shall first discuss issues in relativity and field theory in the long-baseline space quantum experiments in Section \ref{RQFT}. 
Then we analyze quantum teleportation \cite{BB93} and entanglement swapping \cite{ZZ93} in this setting in Secs. \ref{SecQTelep} and \ref{SecEntSwap}, respectively. An introduction to the Bell test is given in 
\ref{Belltest}, and a comment on the ``before-before" and ``after-after" scenarios in the Bell test is given in 
\ref{BBAA}.

\section{Issues in relativity and field theory}
\label{RQFT}

The travel time of a classical light signal from the Earth's surface to the Moon is about 1.3 s, given the mean distance to the Moon from the Earth's center $r^{}_{Moon}=3.85\times 10^8$ m, the mean radius of the Earth $r_{\oplus}=6.37 \times 10^6$ m, and the speed of light $c=3\times 10^8$ m/s \cite{Planets20}. The quantum coherence of Alice's and Bob's photons or atoms should last up to this time scale.
All the instruments of Alice's, Bob's as well as the third or fourth agent (called Charlie and Diana later) should be synchronized with the coordinate time of a bookkeeper within the time resolution needed in this kind of experiments (roughly under 0.1 ps $= 10^{-13}$ s, which is the typical time scale of the driving laser pulse or the wavepackets of photons).

The motion of the Earth, the Moon, the ISS, and the LG are not relativistic, and the Moon's Newtonian gravitational potential experienced by the ISS is roughly $10^{-4}$ of the Earth's (the masses of the Earth and the Moon are $M_{\oplus}= 5.97 \times 10^{24} $ kg and $M^{}_{Moon} = 7.35 \times 10^{22}$ kg, respectively, and the orbital radius of the ISS is $r^{}_A \approx r_{\oplus} + 400\, {\rm km} = 6.77 \times 10^6$ m \cite{Planets20}), while the time scale of our quantum experiment is relatively short.
Thus the ISS could be considered as moving approximately in the Schwarzschild geometry of the Earth,
\begin{eqnarray}
	ds^2 &=& 
	-\left( 1-\frac{2GM_{\oplus}}{c^2 r} \right) c^2 dt^2 + \left( 1-\frac{2GM_{\oplus}}{c^2r} \right)^{-1} dr^2 + \nonumber\\
	& & \hspace{5mm} r^2 \left(d\theta^2 + \sin^2\theta d\varphi^2\right) \label{SchwarzMetric}
\end{eqnarray}
with the Newtonian constant of gravitation $G=6.67\times 10^{-11}$ ${\rm m^3 \, kg^{-1}\, s^{-2}}$. The angular frequency of the ISS is about $d\varphi^{}_A/dt \approx 15.5 \times 2\pi$  ${\rm day}^{-1} = 1.1 \times 10^{-3}$ ${\rm s}^{-1}$. So the proper time of Alice on the ISS is
\begin{equation}
 \tau^{}_{A} \sim  t \sqrt{\left( 1-\frac{2GM_{\oplus}}{c^2r^{}_A} \right) -
    r_A^2 \left(\frac{d\varphi^{}_A}{c \, dt}\right)^2} \approx t \left( 1 - 9.8 \times 10^{-10} \right), \label{tauA}
\end{equation}
implying that the gravitational Doppler effect for Alice is about $\Delta\lambda/\lambda_0 \sim 10^{-9}$ compared with the observer in the orbital axis at radial infinity.

For the LG, since the Moon's and the Earth's Newtonian gravitational potentials experienced by the LG are comparable 
(orbit radius $r^{}_B \sim 10^7$ m and angular velocity $d\varphi^{}_B/dt \sim 2\pi/(7\, {\rm days}) \approx 1 \times 10^{-5}$ ${\rm s}^{-1}$ to the Moon), its NRHO is complicated. Here we only make a very simple, order-of-magnitude estimate by applying (\ref{tauA}) to the LG:
\begin{equation}
  \tau^{}_{B} \sim  t \sqrt{\left[ 1-\frac{2}{c^2}\left(\frac{GM^{}_{Moon}}{r^{}_B} +\frac{GM_{\oplus}}{r^{}_{Moon}}\right) \right] -
     r_{Moon}^2 \left(\frac{d\varphi^{}_{Moon}}{c \, dt}\right)^2}. 
		\label{tauB}
\end{equation}
The result suggests that the gravitational Doppler shift for Bob would be about $\Delta\lambda/\lambda_0 \sim 10^{-11}$, less than the one for Alice. Suppose we synchronize $\tau^{}_{A} = \tau^{}_B=0$ at $t=0$. Then Alice's clock on the ISS would be slower than Bob's clock on the LG by about 1 ns per second read by a bookkeeper in the $z$ axis at infinity.

The speed of the ISS and LG relative to the Earth are of the order of $10^4$ and $10^3$ m/s, respectively (the orbital speed of the ISS is 
about 7.7 km/s, and the mean orbital speed of the Moon is about 1 km/s), and so the relative speed of Alice on the ISS and Bob on the LG would at most be of the order of $10^4$ m/s. In special relativity this can lead to a radial Doppler shift of the order of $\Delta\lambda/\lambda_0 = \sqrt{(1+v/c)/(1-v/c)} \sim 10^{-5}$ \cite{DIn92}, which can dominate over the gravitational Doppler effect in (\ref{tauA}). 
While this would reduce the coherence of the photons from different sources, and so reduce the counting rates of interference experiments such as the Bell state measurement, the radial Doppler effect could be suppressed by executing the experiment in a period when the relative radial motion is negligible.
In this case the transverse Doppler shift will at most be of the order of $\Delta\lambda/\lambda_0 = \sqrt{1/(1-v^2/c^2)} \sim 10^{-10}$, 
which is mainly contributed by the $d\varphi_A/dt$ term in (\ref{tauA}) in our consideration of the gravitational Doppler effect.

For Charlie in our proposed quantum teleportation experiment and Diana in the entanglement swapping experiment placed around the mid-point of Alice and Bob (Sections \ref{SecQTelep} and \ref{SecEntSwap}), if Charlie or Diana takes a transfer vehicle traveling between the ISS and LG, a simple estimate using the Holmann transfer orbit \cite{TM04} suggests that the speed of the vehicle would be about $v = \sqrt{GM_\oplus/r^{}_{1/2}} \approx 1.4\times 10^3$ m/s around the mid-point $r \approx r^{}_{1/2} \equiv (r^{}_A + r^{}_{Moon})/2$ (which is the semi-major axis of that elliptical transfer orbit). This would lead to a radial Doppler shift of the order of $\Delta\lambda/\lambda_0 \approx 5\times 10^{-6}$ even if the ISS and LG have negligible relative radial motion. Again, the efficiency of the Bell state measurement in quantum teleportation or entanglement swapping would be reduced.
Alternatively, suppose Charlie and Diana are carried by a satellite in a high earth orbit [nearly circular orbit at a geocentric distance $O(r^{}_{1/2})$ (e.g. Vela satellites \cite{Vela}), or highly elliptical orbit at an apogee of about $r^{}_{1/2}$ (e.g. IBEX \cite{IBEX})], the radial Doppler shift of the emitted or received photons may be suppressed in those time windows of negligible relative radial motions between the satellite, ISS, and LG.

The Wigner rotation of photon polarization due to the relative motion may be compensated by dynamical methods with the help of polarized classical reference beams traveling with the photons from the source to the receiver \cite{YP17}.
As for the effects arising from the vacuum state of quantum fields, since Alice and Bob's quantum objects are separated far enough while their motions and the environment are complicated, each could be considered as independently interacting with its ambient environment at a finite temperature. As for the (circular) Unruh effect \cite{Unr76, BL83}, 
if Alice's quantum objects on the ISS are coupled to the vacuum state of the field with respect to the Earth, since the ISS has a centripetal acceleration $a \approx 8.7\, {\rm m/s^2}$, the Unruh temperature they may experience will be of the same order of 
$T^{}_U= \hbar a/(2\pi k^{}_B c) \approx 4 \times 10^{-20}$ K, which is much lower than the temperature of their ambient environment.  
While the gravitational potentials ($\sim 1/r$) of the Earth and the Moon experienced by the LG are comparable in value, 
the acceleration ($\sim 1/r^2$) of the LG is dominated by the Moon. Since the NRHO of the LG has the periapsis about 3000 km,  
Bob's quantum objects on the LG would experience 
an acceleration no more than of the order of $G M_{Moon}/(3000\, {\rm km})^2 \approx 0.545 \,{\rm m/s^2}$. This corresponds to an Unruh temperature of about $6\%$ of Alice's at most. Therefore, the Unruh effect is negligible in the present setup.

The size of a single-photon wave packet is usually estimated as the scale of the pulse of the driving laser generating the photons, though a shorter pulse corresponds to a wider spectrum in the frequency or wavelength domain. For example, a pulse of 100 ps $= 10^{-10}$ s corresponds to a photon of 3 cm long \cite{SA88}. The size of the photons are even smaller in recent experiments. They are small enough to be considered as local objects compared with the scale of our proposed nonlocal experiments in outer space. While photons are not really local objects, the locality of the experiments may be ensured operationally by checking the (conditional) probabilities.

\section{Quantum teleportation}
\label{SecQTelep}

The simplest kind of nonlocal quantum experiments is the Bell test, where two local measurements are performed by Alice and Bob on the two quantum objects they hold separately in each process [$M_A$ on $A$, and $M_B$ on $B$ in Figure \ref{NonLocalExpt} (left)]. An example of the Bell test can be found in 
\ref{Belltest}.

The second simplest kind of nonlocal quantum experiments is quantum teleportation \cite{BB93}, which includes at least three quantum objects in each process. One joint measurement is performed on two quantum objects by Alice, in addition to the final measurement of the teleported state on Bob's quantum object to get the (averaged) fidelity of teleportation. The general scenario is shown in Figure \ref{NonLocalExpt} (middle), and described as follows.

Suppose Alice is to send an unknown quantum state carried by some quantum object (photon or atom) $C$ to Bob, but for some reason she cannot deliver $C$ directly to Bob's place. One way to transport this unknown quantum state is to prepare a pair of quantum objects $A$ and $B$ in a known entangled state, and let Alice hold on to $A$ and Bob hold on to $B$. At some moment $t=t_A$ in a reference (bookkeeper's) frame in which Alice and Bob are synchronized, Alice performs a joint measurement $M_A$ on the two objects $C$ and $A$ to project them onto some entangled state, while the quantum state of $B$ at Bob's place is collapsed to some state similar, but not necessarily identical, to the original state of $C$ depending on which entangled state $C$ and $A$ collapse to. Then Alice sends the outcome of the joint measurement $I_A$ to Bob by a classical channel, where the message goes no faster than light. Right after Bob receives this message, he performs a local operation $O_B$ to his quantum object $B$ according to Alice's outcome and the initial entangled state of $A$ and $B$. Then he can reconstruct the unknown state originally carried by $C$ on his quantum object $B$.
An explicit example is given in Section \ref{ExQTelep}.

\begin{figure}
\includegraphics[width=6cm]{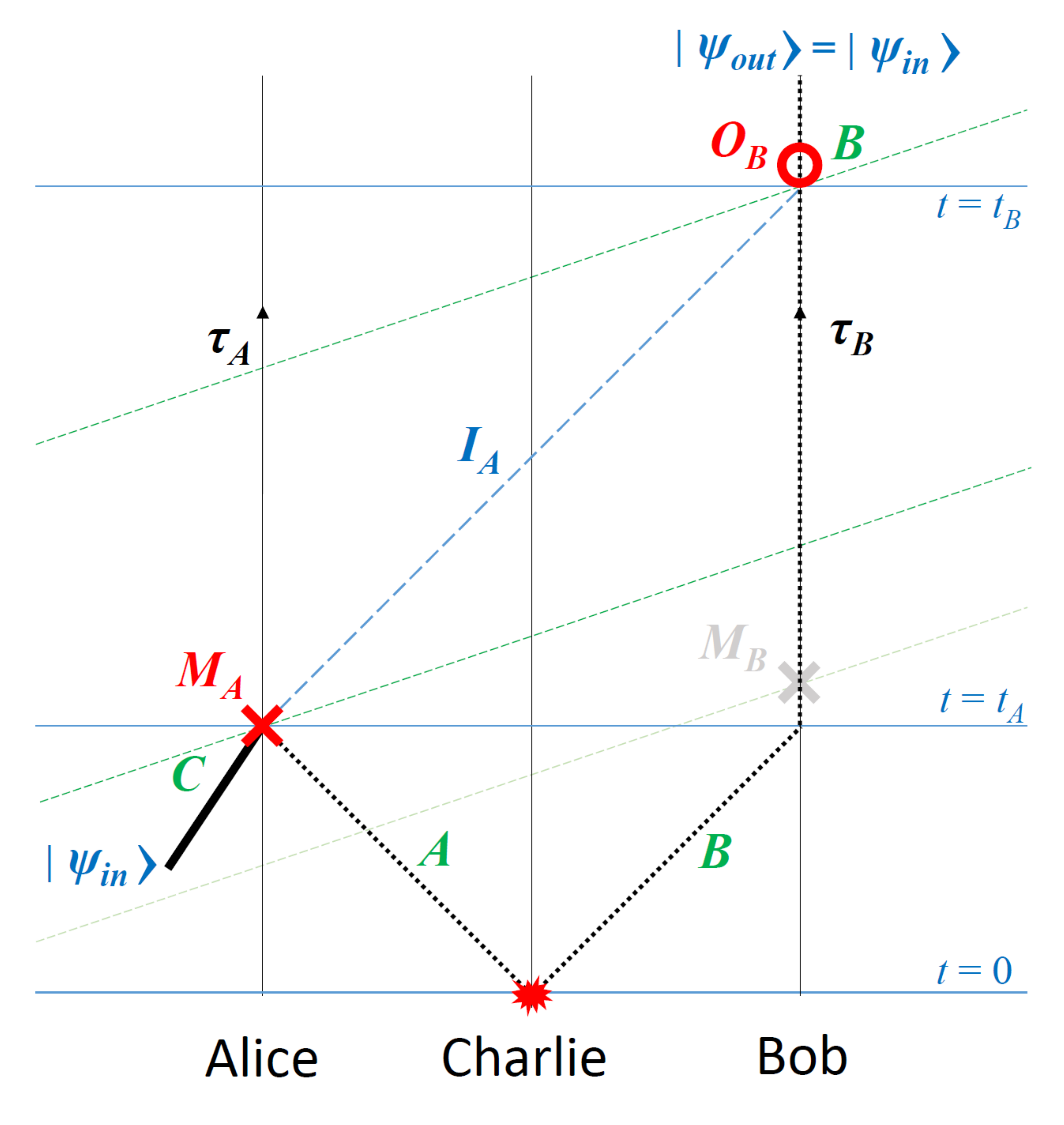}\hspace{.5cm}
\includegraphics[width=6cm]{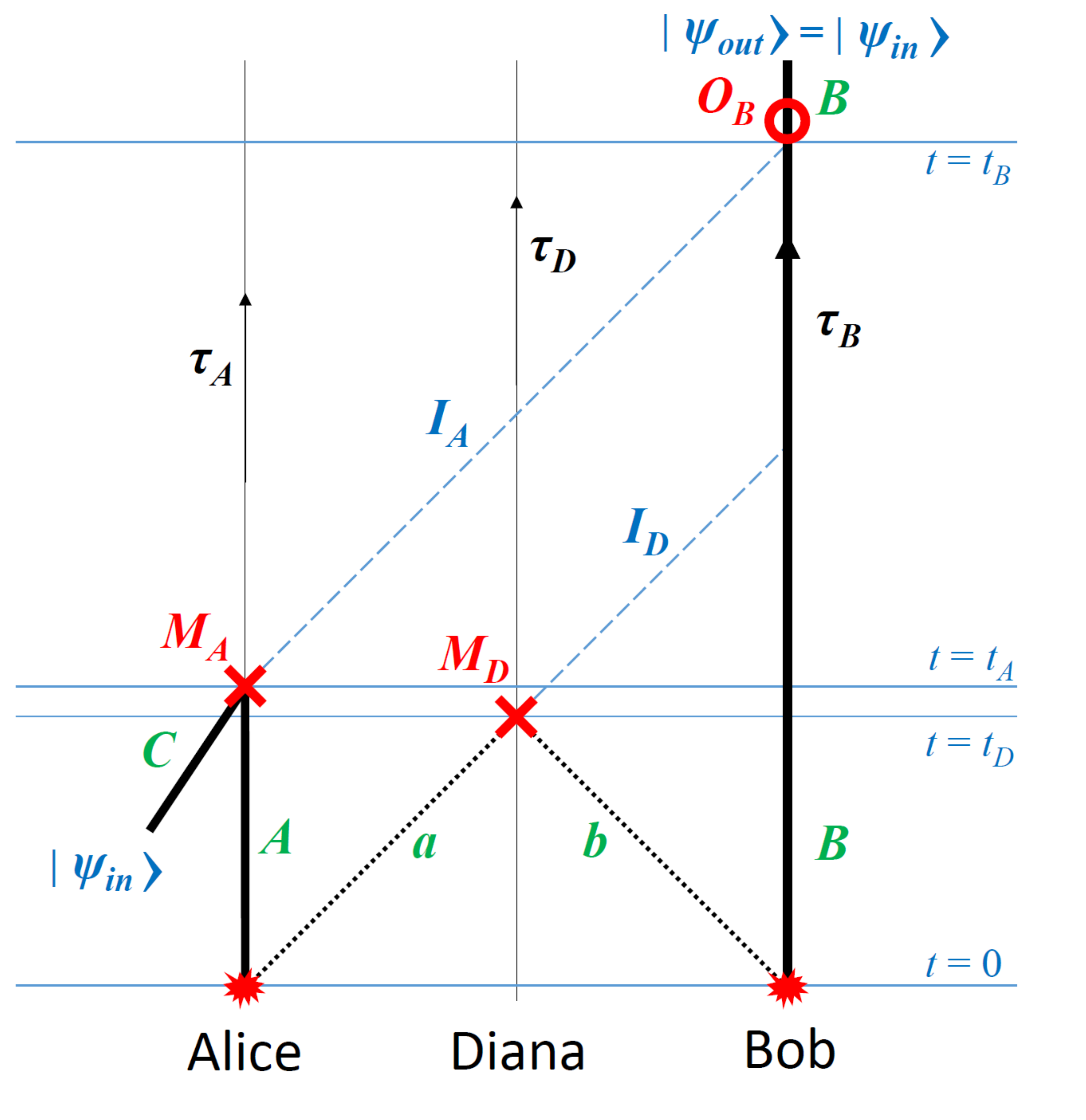}
\caption{Conventional scheme of quantum teleportation (left), and a modified scheme with the qubits ($A$ and $B$) entangled by entanglement swapping (right). Here, the red ``$\circ$" represents local operation events.}
\label{EntSwapQTel}
\end{figure}

Quantum teleportation has been implemented in many physical systems including photons, NMR, trapped ions, atomic ensembles, and solid state systems (for a review, see Ref.\cite{PE15}.) The quantum states used in quantum teleportation are mainly the binary states (qubits) \cite{BB93} and the squeezed coherent states of continuous variables (CV) \cite{Va94}. The entangled pair $A$ and $B$ held by Alice and Bob, respectively, can be distributed by a third agent [Charlie in Figures \ref{EntSwapQTel} (left) and \ref{NonLocalExpt} (middle)], or produced by an entanglement swapping [Figure \ref{EntSwapQTel} (right) and Section \ref{SecEntSwap}.]

In actual implementations in laboratories, the input ``unknown" states are often fed by experimentalists with full knowledge about them, and in some experiments the operation supposed to be carried out by Bob after receiving the classical signal from Alice is never performed physically, but performed virtually in data analysis \cite{RP17} to obtain the fidelity of quantum teleportation [see Eq.(\ref{QTelep01})]. In experiments carried out in this manner (called ``passive" teleportation in Ref.\cite{PE15}) Bob does not need to work hard to keep the quantum coherence of his quantum object $B$ long enough till he receives Alice's classical signals. To obtain the fidelity more efficiently, Bob can perform the measurement early 
[$M_B$ in red in Figure \ref{NonLocalExpt} (middle) and in gray in Figure \ref{EntSwapQTel} (left)], and choose which measurement he would perform on different quantum objects $B$ randomly or following a designed sequence given by the conductor of the experiment. Bob could even perform the measurement on $B$ before Alice's joint measurement on $A$ and $C$ in the bookkeeper coordinates. 

Indeed, in Figure \ref{NonLocalExpt} (middle) one can see that, if Bob's measurement $M_B$ in a passive teleportation experiment is spacelike-separated from Alice's joint measurement $M_A$, then $M_B$ may occur earlier than $M_A$ in the reference frames of some moving bookkeepers (e.g., the green dashed time slices). A description by such a moving bookkeeper is given in Section \ref{ExQTelep}. When Bob is performing $M_B$, he does not know which quantum state that $A$ and $C$ would be collapsed to by Alice, or whether Alice would ever do the joint measurement. A single outcome of Bob's measurement tells nothing. One has to repeat the teleportation many times and collect a series of measurement outcomes from Alice and Bob to obtain the averaged fidelity of quantum teleportation [Figure \ref{EntSwapQTel} (right)], then compare it with
the fidelity of classical teleportation to see the quantum advantage of this process. 

As we mentioned in Section \ref{DSQLadv}, in our proposed experiment at a very long baseline, Alice's measurement events in a series of the repeated quantum teleportation can be all causally independent of Bob's and so quantum nonlocality in our result would be much cleaner than the existing experiments [Figure \ref{BellQTel} (right)]. Nevertheless, the quantum advantage in teleportation may not be conclusive in such a high-loss QI experiment, which suffers from problems similar to the detector-efficiency loophole and the coincidence-time loophole in the Bell test \cite{La98, LG04}. We will discuss this in Section \ref{Floophole}.
 

\subsection{Quantum teleportation with photon qubits}
\label{ExQTelep}

In the original scenario of quantum teleportation proposed in Ref.\cite{BB93}, the quantum objects $A$, $B$, and $C$ carry binary states 0 and 1 as qubits. We shall use it as a simple example to illustrate the implementation of quantum teleportation.

Suppose the quantum state $\alpha |0\rangle^{}_C + \beta |1 \rangle^{}_C $ with the unknown complex constants $\alpha$ and $\beta$ ($|\alpha|^2+|\beta|^2=1$) is carried by photon $C$ held by Alice. At $t=0$, Charlie prepares photons $A$ and $B$ in a Bell state, e.g., $(|0\rangle^{}_A |1\rangle^{}_B -|1\rangle^{}_A |0\rangle^{}_B )/\sqrt{2}$, so the initial state of the three photons $A$, $B$, and $C$ is
\begin{eqnarray}
& & | \psi_0 \rangle = \frac{1}{\sqrt{2}}\left(\frac{}{}|0\rangle^{}_A |1\rangle^{}_B - |1\rangle^{}_A |0\rangle^{}_B\right) 
	\otimes \left( \frac{}{} \alpha |0\rangle^{}_C + \beta |1 \rangle^{}_C \right)  \nonumber\\
&=& \frac{1}{2}\left[ |{\cal B}_{00}\rangle^{}_{AC} \otimes \left( \frac{}{}\alpha |1\rangle^{}_B - \beta |0 \rangle^{}_B \right)+
	 |{\cal B}_{01}\rangle^{}_{AC} \otimes \left( \frac{}{}-\alpha |0\rangle^{}_B + \beta |1 \rangle^{}_B \right)\right. \nonumber\\ & &
	 \left. + |{\cal B}_{10}\rangle^{}_{AC} \otimes \left( \frac{}{}\alpha |1\rangle^{}_B + \beta |0 \rangle^{}_B \right)+
	 |{\cal B}_{11}\rangle^{}_{AC} \otimes \left( \frac{}{}\alpha |0\rangle^{}_B + \beta |1 \rangle^{}_B \right)  \right]
\label{QTelep01}
\end{eqnarray}
where the four orthonormal Bell states are defined as
\begin{equation}
   |{\cal B}_{xy}\rangle \equiv \frac{1}{\sqrt{2}}\left(\frac{}{}|0y\rangle + (-1)^x |1\bar{y}\rangle\right),
\end{equation}
where $x,y=0,1$ are binary digits.
Then Charlie sends photon $A$ to Alice and $B$ to Bob separated in space. 
When Alice receives photon $A$ from Charlie, she makes a Bell measurement [$M_A$ in Figure \ref{EntSwapQTel} (left)] jointly on photons $A$ and $C$, which are projected to one of the four Bell states $|{\cal B}_{xy}\rangle_{AC}$ after the measurement. At the same time in the bookkeeper coordinates, from (\ref{QTelep01}), photon $B$ at Bob's place collapses to the state $\alpha |\bar{y}\rangle^{}_B -(-1)^x \beta |y \rangle^{}_B$, depending on the outcome $xy$ of Alice's measurement. Once Bob receives the outcome $xy$ from Alice, Bob can accordingly make a local operation $\hat{\sigma}^B_{\bar{x},x\oplus y}$ on $B$ [$O^{}_B$ in Figure \ref{EntSwapQTel}], with $\hat{\sigma}^B_\mu \equiv \{\hat{I}_{}^B, \hat{\sigma}_x^B, \hat{\sigma}_y^B, \hat{\sigma}_z^B\}$, $\mu=00,01,10,11$. This will turn the state of photon $B$ to $\frac{}{}\alpha |0\rangle^{}_B + \beta |1 \rangle^{}_B$, which is the original state carried by $C$, without knowing the values of $\alpha$ or $\beta$.
For example, if Alice finds that photons $A$ and $C$ are in the state $|{\cal B}_{11}\rangle^{}_{AC}$, then $|\psi_0\rangle$ must have been collapsed to 
\begin{equation}
  |{\cal B}_{11}\rangle^{}_{AC} \otimes \left( \frac{}{}\alpha |0\rangle^{}_B + \beta |1 \rangle^{}_B \right),
\end{equation}
and the local operation by Bob should be $\hat{\sigma}^B_{\bar{1},1\oplus 1} = \hat{\sigma}^B_{00} =\hat{I}^B$.

In laboratories, the input state initially on photon $C$ is produced by experimentalists and the values of $\alpha$ and $\beta$ may actually be known, and $|{\cal B}_{11}\rangle^{}_{AC}$ can be measured straightforwardly in the coincidence detection \cite{We94, BM95}.\footnote{So far the Bell efficiencies \cite{PE15} of the experiments applying the photon-polarization qubits are no more than 50\% since currently only two of the four Bell states can be distinguished in a single quantum optical measurement process.}
Also, instead of making a local operation before measurement, Bob may directly perform a projective measurement of $\hat{\sigma}^B_\mu$ on photon $B$ without carrying out any local operation \cite{PE15, RP17} to specify the final state of photon $B$ 
(passive teleportation \cite{PE15}).
To do so one needs to prepare an ensemble of three-photon sets ($A$, $B$, $C$), with each three-photon set carrying the same state $|\psi_0\rangle$ in (\ref{QTelep01}). Then, after a series of the same processes, the bookkeeper can select the data with Alice's outcome being $|{\cal B}_{11}\rangle^{}_{AC}$ to estimate the averaged fidelity of quantum teleportation using the associated Bob's outcomes.

In passive teleportation Bob's measurements can be made much earlier than when Bob receives Alice's classical signals. In other words, the events of Bob's measurements can be spacelike separated from Alice's. This implies that, in some moving bookkeeper's coordinates, the joint measurement by Alice on photons $A$ and $C$ may occur later than the Bob's measurement on photon $B$ in the coordinate time (e.g., the green $t'$-slices in Figure \ref{NonLocalExpt}), and the history described by the observers in those coordinates will be different from the conventional one described above. In this scheme, when Bob is measuring photon $B$'s, he does not know what the outcomes of Alice's joint measurements would be. So his strategy may be to randomly choose to measure one of the direction $\hat{\sigma}^B_\mu$ in the Bloch sphere for each photon $B$. For example, if Bob chooses to measure $\hat{\sigma}_{11}^B = \hat{\sigma}_{z}^B$ (whose eigenstates are $|0 \rangle^{}_B$ and $|1 \rangle^{}_B$), then $|\psi_0\rangle$ will have a probability of $1/2$ to be collapsed to
\begin{equation}
	\frac{1}{2}\left[ \frac{}{} \alpha\left(\frac{}{}|{\cal B}_{11}\rangle^{}_{AC} -|{\cal B}_{01}\rangle^{}_{AC} \right) 
	+\beta\left(\frac{}{} |{\cal B}_{10}\rangle^{}_{AC}-|{\cal B}_{00}\rangle^{}_{AC}\right) \frac{}{} \right]\otimes|0 \rangle^{}_B,
\label{PMSBob0}
\end{equation}
which has a probability of $|\alpha/2|^2$ to be further collapsed to $|{\cal B}_{11}\rangle^{}_{AC}\otimes|0 \rangle^{}_B$ by Alice, and the other $1/2$ to be collapsed to
\begin{equation}
	\frac{1}{2}\left[ \frac{}{} \alpha\left(\frac{}{}|{\cal B}_{00}\rangle^{}_{AC} +|{\cal B}_{10}\rangle^{}_{AC} \right) 
	+\beta\left(\frac{}{} |{\cal B}_{11}\rangle^{}_{AC}+|{\cal B}_{01}\rangle^{}_{AC}\right) \frac{}{} \right]\otimes |1 \rangle^{}_B ,
\label{PMSBob1}
\end{equation}
which has a probability of $|\beta/2|^2$ to be further collapsed to $|{\cal B}_{11}\rangle^{}_{AC}\otimes|0 \rangle^{}_B$.
After Alice performs the Bell measurements on photons $A$ and $C$, the bookkeeper collects all the data of the outcomes from Alice and Bob via classical channel, then selects the data with Alice's outcomes being $|{\cal B}_{11}\rangle^{}_{AC}$ to see how closely the (conditional) expectation values satisfy
\begin{eqnarray}
  \frac{{}^{}_{{}_{AC}}\hspace{-.8mm}\langle {\cal B}_{11} |{\rm Tr}[ \hat{\sigma}^{B}_\mu \hat{\rho}_0]| 
	      {\cal B}_{11}\rangle\hspace{-.8mm}{}^{}_{{}_{AC}}}
	  {{}^{}_{{}_{AC}}\hspace{-.8mm}\langle {\cal B}_{11} |{\rm Tr}\hat{\rho}_0| {\cal B}_{11}\rangle\hspace{-.8mm}{}^{}_{{}_{AC}}} &=& 
		\left(\frac{}{}{}^{}_{{}_{C}}\hspace{-.9mm}\langle 0 | \alpha^* +
		  {}^{}_{{}_{C}}\hspace{-.9mm}\langle 1|\beta^* \right) \hat{\sigma}^C_\mu 
		\left(\frac{}{} \alpha\, |0\rangle\hspace{-.8mm}{}^{}_{{}_{C}}+\beta \, |1\rangle\hspace{-.8mm}{}^{}_{{}_{C}} \right) 
		\\&\equiv&
		\langle \hat{\sigma}^C_\mu\rangle\hspace{-.8mm}{}^{}_{{}_{C}}\nonumber
\label{fideavg}
\end{eqnarray}
for all $\mu=01,10,11$ (the case of $\mu=00$ is trivial), as can be derived straightforwardly from (\ref{QTelep01}) with $\langle \hat{\sigma}^C_\mu\rangle\hspace{-.8mm}{}^{}_{{}_{C}} = (1, 2{\rm Re}[\alpha\beta^*], 
-2{\rm Im}[\alpha\beta^*], |\alpha|^2-|\beta|^2)$ and $\hat{\rho}_0 \equiv | \psi_0 \rangle \langle \psi_0 |$.
When the averaged fidelity is $1$, the experimental results for the left-hand side of (\ref{fideavg}) will have identical values to the right-hand side.  

While the quantum states in the above scheme for a moving bookkeeper are defined on different time-slices and look different from those for the bookkeeper at rest in the conventional scheme, canonical quantum mechanics guarantee that all the measurement outcomes and the expectation values of physical observables in the above described history are consistent with those in the conventional history, 
once the initial states in different coordinates are defined on the same fiducial time-slice where they are identical up to coordinate and gauge transformations \cite{LCH15,Lin11}.

\subsection{Coincidence-time loophole in the fidelity of teleportation}
\label{Floophole}

With the high loss of photons in our very long baseline experiment, the detector-efficiency and coincidence-time loopholes in the Bell test may be significant. In similar conditions the averaged fidelity of quantum teleportation of binary states can also exceed the ideal ``classical" bound.

The fidelity of teleportation from Alice to Bob with the in-state $|\psi^{}_{\rm in}\rangle = \alpha |0\rangle_{C} + \beta |1\rangle_{C} \equiv \cos (\theta_{_C}/2)|0\rangle_{C} + e^{i\varphi_{_C}}\sin (\theta_{_C}/2) |1\rangle_{C} $ up to a global phase before Alice's Bell-state measurement and the out-state $|\psi^{}_{\rm out}\rangle$ after Bob's operation is defined as 
${\cal F} = \left| \langle \psi^{}_{\rm out} | \psi^{}_{\rm in} \rangle \right|^2$, which can be written as
\begin{equation}
  {\cal F} = {\rm Tr}\, \hat{\rho}^{}_{\rm in}\hat{\rho}^{}_{\rm out} = 
	\frac{1}{2}\left( 1+ {\vec r}^{}_{\rm in}\cdot {\vec r}^{}_{\rm out}\right) \label{FiQT}
\end{equation}
with the density matrices of the in-state $\hat{\rho}_{\rm in} = (\hat{1}+{\vec r}_{\rm in}\cdot \hat{\vec\sigma}
)/2$ and out-state $\hat{\rho}_{\rm out} = (\hat{1}+{\vec r}_{\rm out}\cdot \hat{\vec\sigma}
)/2$. Here, we denote $\hat{\vec\sigma}
\equiv \hat{\sigma}_x {\vec x} +\hat{\sigma}_y {\vec y}+\hat{\sigma}_z {\vec z}$, and
\begin{eqnarray}
  {\vec r}^{}_{\rm in} &\equiv& \langle \hat{\sigma}_x\rangle^{}_{\rm in}\, {\vec x} + \langle\hat{\sigma}_y\rangle^{}_{\rm in}\, {\vec y} + 
	\langle\hat{\sigma}_z\rangle^{}_{\rm in}\, {\vec z} \nonumber\\ &\equiv&
   {\vec x}\,\cos \varphi_{_C} \sin\theta_{_C} + {\vec y}\,\sin \varphi_{_C} \sin\theta_{_C} + {\vec z}\,\cos \theta_{_C} 
	\label{rin}
\end{eqnarray}
with $\{{\vec x},{\vec y},{\vec z}\}$ the orthonormal basis of the ${\bf R}^3$ that the Bloch sphere is embedded in, and the expectation values of the Pauli matrices $\langle \hat{\sigma}_j \rangle^{}_{\rm in} \equiv \langle \psi^{}_{\rm in}| \hat{\sigma}_j |\psi^{}_{\rm in}\rangle$, $j=x,y,z$, with respect to the in-state. The definition of ${\vec r}^{}_{\rm out}$ for the out-state is similar. Below, we will keep using (\ref{FiQT}) as the definition of fidelity for each run of teleportation even if the out-state is not a pure state ($|{\vec r}^{}_{\rm out}|^2 < 1$).

Suppose Charlie produces a series of photon pairs. Each photon pair is in a separable but still ``classically" correlated state 
\begin{equation}
  \hspace{-2mm}
  \left[\cos \frac{\tilde\theta}{2}\,|0\rangle_{A}^{} + e^{i\tilde\varphi}\sin \frac{\tilde\theta}{2}\, |1\rangle_{A}^{} \right] \otimes 
  \left[\cos \frac{\pi-\tilde\theta}{2}\,|0\rangle_{B}^{} + e^{i(\pi+\tilde\varphi)}\sin \frac{\pi-\tilde\theta}{2}\, |1\rangle_{B}^{}\right]
	\label{Clphotons}
\end{equation}
with the state-vectors of photon $A$ and photon $B$ in the same pair always pointing to opposite directions on the Bloch sphere (namely, the two photons' polarizations are perpendicular to each other). The directions of the state-vectors for different photon-pairs are prepared to be evenly distributed over the Bloch sphere ($\tilde{\varphi}\in [0,2\pi]$ and $\tilde{\theta}\in [0,\pi]$). 
From Section \ref{ExQTelep}, after the Bell-state measurement by Alice, the probability that photons A and C are collapsed to the Bell state $|{\cal B}_{xy}\rangle_{_{AC}}$ is
\begin{eqnarray}
  {\rm Pr}\left({\cal B}_{xy}\right) &=& \frac{1}{4}\left\{ 1 +(-1)^y \cos \left( \theta_{_C} + (-1)^y\tilde{\theta}\right) +\right. 
	\nonumber\\ & & \hspace{5mm}\left.
	\left[1+(-1)^x\cos \left(\varphi_{_C} + \tilde{\varphi}\right) \right] \sin\theta_{_C}\, \sin\tilde{\theta}  \right\} .
\end{eqnarray}
Then, after the operation by Bob on each photon pair, the fidelity of teleportation for each run may depend on  the directions of the in-state $(\theta_{_C}, \varphi_{_C})$ and the photon pair $(\tilde{\theta},\tilde{\varphi})$, while the dependence of Alice's outcome $(x,y)$ 
is obliterated by Bob's operation.  

After many runs of teleportation, the ``classical" averaged fidelity of teleportation will be given by
\begin{equation}
  {\cal F}_{av} \equiv \frac{1}{4\pi}\int_0^\pi \sin \tilde{\theta} d\tilde{\theta} \int_0^{2\pi} d\tilde{\varphi}\, 
	  {\cal F}(\theta_{_C}, \varphi_{_C}; \tilde{\theta},\tilde{\varphi}) = 
		\frac{1}{2}\left( 1+ {\vec r}^{}_{\rm in}\cdot \langle\hspace{-1mm}\langle{\vec r}^{}_{\rm out}\rangle\hspace{-1mm}\rangle\right) 
	\label{FiQTav}
\end{equation} 
with the ensemble average of ${\vec r}^{}_{\rm out}$, 
\begin{eqnarray}
\langle\hspace{-1mm}\langle{\vec r}^{}_{\rm out}\rangle\hspace{-1mm}\rangle  
&\equiv&\sum_{x,y=0,1}
	 \frac{1}{4\pi}\int_0^\pi \sin \tilde{\theta} d\tilde{\theta} \int_0^{2\pi} d\tilde{\varphi}\, 
	{\rm Pr}\left({\cal B}_{xy}\right) \times \nonumber\\
& &\hspace{5mm}\left[ {\vec x}\, (-1)^x_{} \cos\tilde{\varphi}\sin\tilde{\theta} - 
	{\vec y}\, (-1)^{x\oplus y}_{} \sin\tilde{\varphi}\sin\tilde{\theta} + {\vec z}\,(-1)^y_{}\cos\tilde{\theta}\,\right] \nonumber\\
&=& 4\left(\frac{{\vec x}}{12}\cos\varphi_{_C}\sin\theta_{_C} + \frac{{\vec y}}{12} \sin\varphi_{_C}\sin\theta_{_C} 
	  + \frac{{\vec z}}{12}\cos\theta_{_C}\right) \nonumber\\
&=& \frac{1}{3} {\vec r}_{\rm in}
\end{eqnarray}
after some algebra. Then we recover the ``classical" upper bound of the averaged fidelity of teleportation ${\cal F}_{av}={\cal F}_{cl} \equiv 2/3$ \cite{MP95}, though here the expectation values $\langle \hat{\sigma}_k\rangle_{\rm out}$ in ${\vec r}_{\rm out}$ have been evaluated quantum-mechanically using the out-states.

If there exists some dependence between the input and the output photons {\it locally} introduced by the optical devices, the classical correlation of the above separated photon pair may be converted to the correlation in the measurement outcomes of Alice and Bob, which would easily make the averaged fidelity exceed the classical upper bound if there is any statistical bias in data analysis. To verify this, we have done a computer simulation using (\ref{Clphotons}) without considering entanglement or quantum nonlocality in the photon pairs but simply tagging each local photon detection a triggering time which is randomly distributed in a period proportional to the square of the angular difference between the in- and out-photon polarizations \cite{DR05, DR07a, DR07b}. Indeed, we can make the averaged fidelity of teleportation greater than the ideal classical bound ${\cal F}_{cl}=2/3$ by setting a sufficiently narrow time-window in selecting the coincidence detection events. 

\section{Entanglement swapping}
\label{SecEntSwap}

Entanglement swapping is an immediate generalization of quantum teleportation from three participating quantum objects to four. The scenario is as follows. Suppose Alice, Bob, and Diana are separated in space with Diana situated at the mid-point between Alice and Bob, and all of them have synchronized their clocks to the time coordinate of some bookkeeper's frame [see Figure \ref{NonLocalExpt} (right), in which the blue horizontal lines represent the time slices in this frame.]
At $t=0$ Alice produces an entangled photon pair $(A,a)$, and sends photon $a$ to Diana while keeping photon $A$. At the same moment in the bookkeeper's coordinates, Bob also produces an entangled photon pair $(B,b)$ in a similar state, and sends photon $b$ to Diana but keeps photon $B$.  
Initially $A$ and $B$ are not entangled, and there is no entanglement between $a$ and $b$, either.
When Diana receives photons $a$ and $b$ from Alice and Bob, respectively, she performs a joint measurement $M_D$ on these two photons.  Then, the quantum state of the photons will be projected to a direct product of an entangled state of $a$ and $b$ and another entangled state of $A$ and $B$.
This process is known as ``entanglement swapping" \cite{ZZ93} (for an explicit example, see Section \ref{ExEntSwap}.) The created entanglement of $A$ and $B$ can be further verified by measuring some entanglement witness \cite{HZ15}, or by the Bell test for binary states [see 
\ref{Belltest}.] 
Either way, an ensemble average of a collection of the measurement outcomes is needed.

\begin{figure}
\includegraphics[width=7cm]{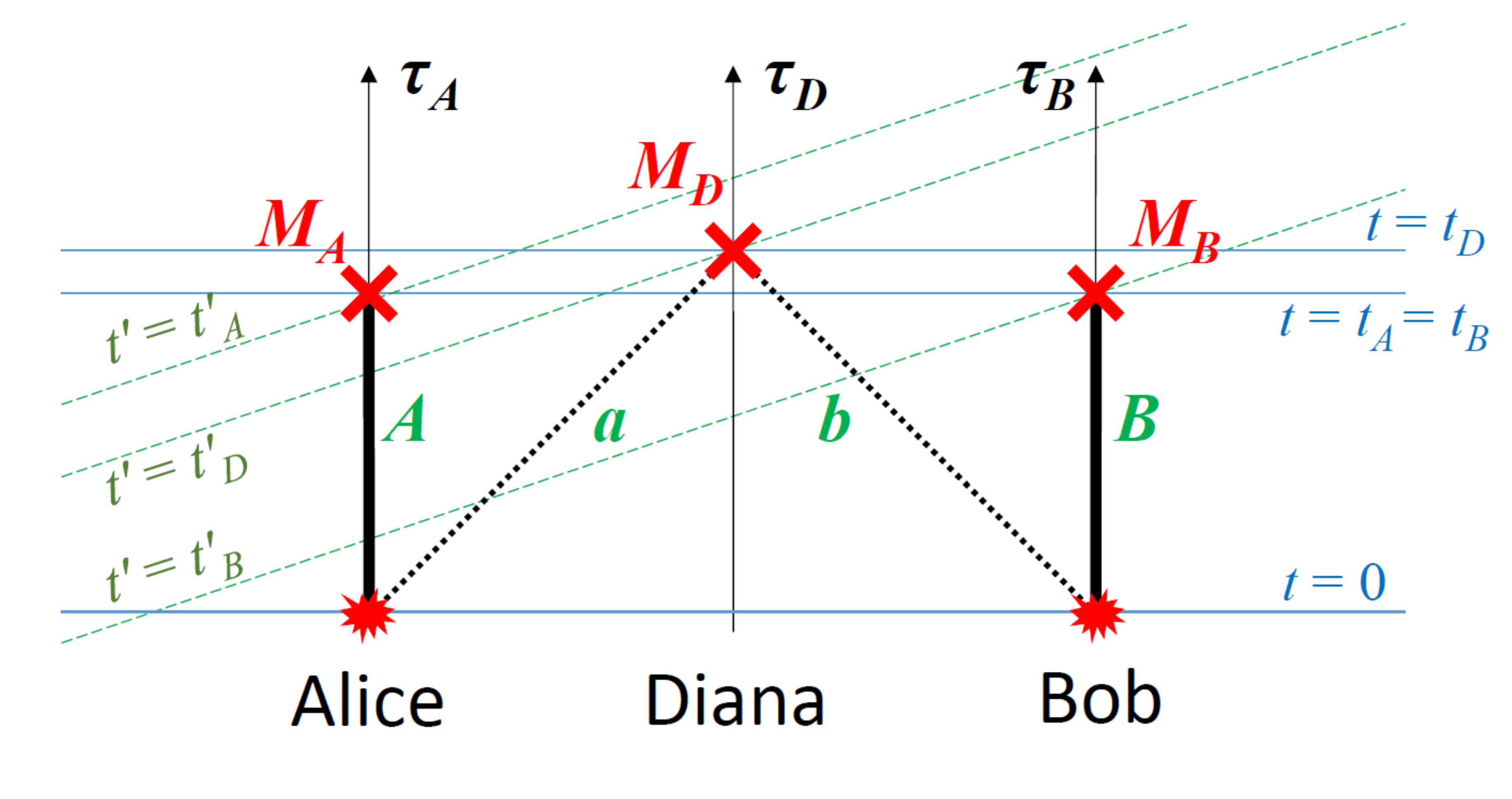}\hspace{.2cm}
\includegraphics[width=6cm]{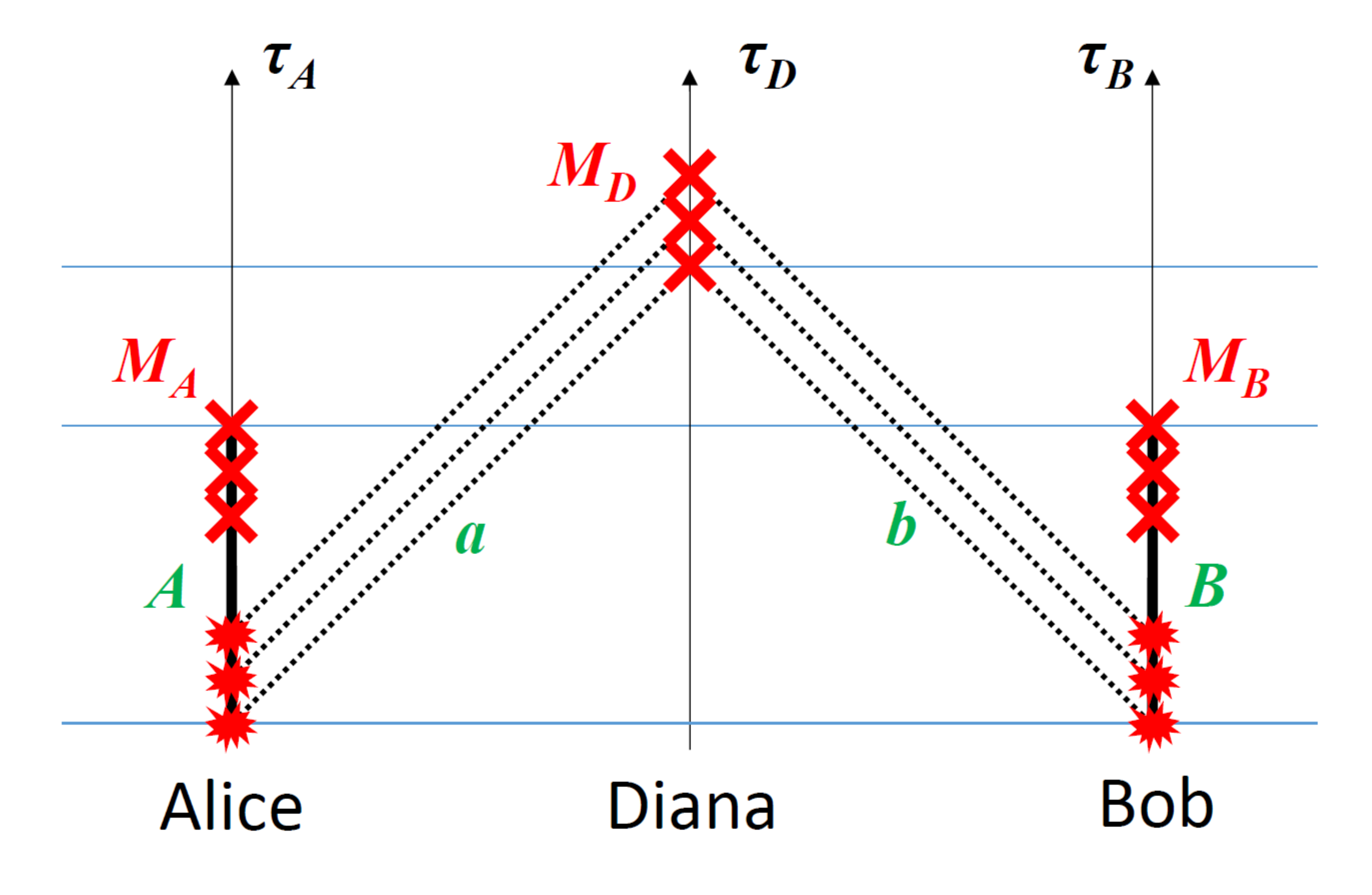}
\caption{The left plot is a tilt of the entanglement swapping from Figure \ref{NonLocalExpt} (right).
Between the ISS and LG a likely scheme is shown in the right plot. Here $t_D > t_A, t_B$.}
\label{EntSwapDLater}
\end{figure}


Right after (in the bookkeeper's coordinates) Diana's joint measurement ($t>t_D$), Alice and Bob separately measures photons $A$ and $B$ [$M_A$ at $t=t_A$ and $M_B$ at $t=t_B=t_A$ in Figure \ref{NonLocalExpt} (right)], respectively, following the planning 
of the experiment. 
The bookkeeper then collects the measurement outcomes from Alice, Bob, and Diana for a series of the photon double pairs ($A,a; B,b$), and average those sets for the individual double pair $(A, a; B, b)$ with Diana's measurement outcome being entangled $(a, b)$.
The bookkeeper then checks if the CHSH inequality of $(A,B)$ is violated as in (\ref{CHSHBell}) or the entanglement witness of $(A, B)$ is evident.
If it is, in the bookkeeper's description the entanglement between Alice and Bob's photons $A$ and $B$ has been created by Diana's measurement at $t=t_D$. 
Note that at this moment there is not yet any direct (retarded) interaction or classical information exchange between photons $A$ and $B$ after they are created at $t=0$. 

Indeed, 
since the quantum state of the combined system at each specific moment is defined on the whole time slice,
the entanglement of photons $A$ and $B$ and the outcome of Diana's measurement occur simultaneously in the whole space at $t=t_D$ in the bookkeeper's coordinates.
For the three measurement events $M_A$, $M_B$, and $M_D$ which are spacelike separated, as shown in Figure \ref{NonLocalExpt} (right), when Alice is performing $M_A$, she is still outside the future lightcone of the event at $t=0$ that Bob produces entangled photon pair $B$ and $b$ ($t_D 
< t_A < 2t_D$), and so she does not know the existence of Bob's photon pair $(B, b)$ at that time. 
Bob does not know the existence of Alice's photon pair $(A, a)$ when performing $M_B$ at $t=t_B$ ($t_D 
< t_B < 2t_D$), either. Moreover, when Alice and Bob are performing their local measurements, they have no idea about the outcome of Diana,  they don't even  know whether Diana has performed $M_D$ or not. The created quantum entanglement between photons $A$ and $B$ will not be realized by anyone until the bookkeeper, Alice, or Bob receives all the outcomes (of $M_A$, $M_B$, and $M_D$), performs the averages, and compares. 

The situation becomes more intriguing if $M_A$ and $M_B$ both occur earlier than $M_D$ in the bookkeeper's frame, while the events $M_A$, $M_B$, and $M_D$ are still all spacelike separated [Figure \ref{EntSwapDLater} (left), where $0 < t_A = t_B < t_D$.] In this case, the final result is the same as the above conventional ones but the description of the intermediate states are different \cite{Pe00}. 
The bookkeeper's description may turn out to be the following:
The entanglement between $a$ and $b$ at $t=t_A=t_B$ is created by Diana's measurement later at $t=t_D > t_A$ and so the entanglement distribution can go backward in the coordinate time from $M_D$ to $M_A$ and $M_B$, once the events $M_A$ and $M_B$ are spacelike separated from $M_D$.

The above description would make more sense if the spacetime region where all the measurements by Alice occur is spacelike separated from the spacetime regions of Bob's measurements and  Diana's, as in Figure \ref{EntSwapDLater} (right). 
In this way Alice and Bob's measurement outcomes entering into the statistics for verifying the entanglement between $A$ and $B$ will have a clean causal independence. 

This scenario is not so weird when one only looks at the outcomes of the measurements, rather than the quantum states. 
In Section \ref{ExEntSwap} we offer a formal description that the bookkeeper would provide.
Only the data of the outcomes of $M_A$ and $M_B$ which are post-selected according to the detectable outcomes of $M_D$ can give the expectation values showing that $A$ and $B$ are entangled. 
Without the reference of the outcomes of $M_D$, one cannot see the evidence that $A$ and $B$ are entangled: 
the statistics of all the outcomes of Alice and Bob's local measurements would simply respect the CHSH inequality.



\subsection{Entanglement swapping of photon qubits}
\label{ExEntSwap}

Suppose a series of entangled photon pairs $(A, a)$ in the state $\left(|0\rangle^{}_A |1\rangle^{}_a + |1\rangle^{}_A |0\rangle^{}_a \right)/\sqrt{2}$ are produced by Alice in a period much shorter than the travel time from Alice to Diana.
Right after each photon pair is produced Alice sends photon $a$ to Diana and keeps photon $A$. The emitted photon $a$'s form a train with the total length much shorter than the distance between Alice and Diana. 
At each moment in the bookkeeper's coordinates when Alice produces an entangled photon pair, Bob also produces an entangled photon pair $(B,b)$ in the state $\left(|0\rangle^{}_B |1\rangle^{}_b + |1\rangle^{}_B |0\rangle^{}_b \right)/\sqrt{2}$, and sends photon $b$ to Diana but keeps photon $B$ [see Figure \ref{EntSwapDLater} (right)].

Right after the trains of photons $a$ and $b$ depart, the quantum state of each photon double pair produced by Alice and Bob at the same moment in the bookkeeper's frame is
\begin{equation}
  | \psi_{0}\rangle =
	\frac{1}{\sqrt{2}}\left( \frac{}{} |0\rangle^{}_A |1\rangle^{}_a + |1\rangle^{}_A |0\rangle^{}_a \right) \otimes
	\frac{1}{\sqrt{2}}\left( \frac{}{} |0\rangle^{}_B |1\rangle^{}_b + |1\rangle^{}_B |0\rangle^{}_b \right).
\label{Psi0}
\end{equation}
When Diana receives photons $a$ and $b$ from Alice and Bob, respectively, she performs a joint Bell-state measurement $\hat{M}_D$ on these two photons. For example, a straightforward coincidence detection in the output of a Hong-Ou-Mandel interferometer \cite{HOM87} for photons $a$ and $b$ implies a projection \cite{We94, BM95}
\begin{equation}
  \hat{M}_{D_{11}} = \frac{1}{\sqrt{2}}\left( \frac{}{} |0\rangle^{}_a |1\rangle^{}_b -|1\rangle^{}_a |0\rangle^{}_b \right)
	\left( \frac{}{} {}^{}_a\langle 0 | {}^{}_b\langle 1 | -{}^{}_a\langle 1 | {}^{}_b\langle 0 |\right)\frac{1}{\sqrt{2}}
\end{equation}
to the Bell state $|\Psi^{(-)}\rangle \equiv |{\cal B}_{11}\rangle_{ab}$.
After Diana's joint measurement on $a$ and $b$, the quantum state of the whole system is projected to
\begin{eqnarray}
  |\psi_{1}\rangle &=& 
	\frac{\hat{M}_{D_{11}}| \psi_{0}\rangle}{\sqrt{\langle\psi_{0}|\hat{M}^\dagger_{D_{11}}\hat{M}^{}_{D_{11}}| \psi_{0}\rangle}}
	\nonumber\\ &=& 
	\frac{1}{\sqrt{2}}\left( \frac{}{} |0\rangle^{}_a |1\rangle^{}_b -|1\rangle^{}_a |0\rangle^{}_b \right) \otimes
	\frac{1}{\sqrt{2}}\left( \frac{}{} |1\rangle^{}_A |0\rangle^{}_B -|0\rangle^{}_A |1\rangle^{}_B \right) \nonumber\\ &=&  
	-|{\cal B}_{11}\rangle_{ab}^{} \otimes |{\cal B}_{11}\rangle_{AB}^{}, 
\label{Psi1}
\end{eqnarray}
where photons $A$ and $B$ are entangled.
Right after (in the bookkeeper's coordinates) Diana's joint measurement ($t>t_D$), 
Alice and Bob can perform the Bell test given in 
\ref{Belltest} to check the entanglement of $A$ and $B$.
The bookkeeper then collects the measurement outcomes from Alice, Bob, and Diana on the whole series of the photon double pairs ($A,a; B,b$), and average those sets for the individual double pair of photons with Diana's measurement outcome being $|{\cal B}_{11}\rangle_{ab}$ to see if the CHSH inequality is violated as in (\ref{CHSHBell}).

\subsection{Delayed-choice entanglement swapping}
\label{DelayedCh}

If both the Alice and Bob's measurements $M_A$ and $M_B$ take place before Diana's Bell measurement $M_D$ in bookkeeper coordinates while three measurement events are spacelike separated \cite{Pe00, JZ02}, the bookkeeper's description goes as follows. First, in $M_A$ ($M_B$) Alice (Bob) freely chooses to measure the property $Q^{}_A$ or $P^{}_A$ ($Q^{}_B$ or $P^{}_B$) of each photon in the series of photons $A$ ($B$) locally. 
The associated photon $a$'s ($b$'s) state is changed accordingly and disentangled from $A$ ($B$) after the projections at $t=t_A=t_B$. 
Thus Diana is measuring the reduced state of photons $a$ and $b$,
\begin{equation}
  \hat{\rho}_1 = {\rm Tr}_{AB}\, \hat{\rho}_0 = \frac{1}{4} \sum_{x,y=0,1}|{\cal B}_{xy}\rangle^{}_{ab}\langle{\cal B}_{xy}|
\end{equation}
with $\hat{\rho}_0 = |\psi_0\rangle \langle \psi_0|$ from (\ref{Psi0}).
If the outcome of $M_D$ is $|{\cal B}_{11}\rangle_{ab}$, then the expectation value in (\ref{CHSHBell}) can be obtained by calculating
\begin{equation}
  \frac{ ^{}_{ab}\langle {\cal B}_{11}| {\rm Tr}_{AB}\left[(Q^{}_A Q^{}_B + P^{}_A Q^{}_B + P^{}_A P^{}_B - Q^{}_A P^{}_B)\hat{\rho}_0\right] | {\cal B}_{11}\rangle^{}_{ab}}
	  {^{}_{ab}\langle {\cal B}_{11}| \hat{\rho}_1 | {\cal B}_{11}\rangle^{}_{ab}},
\end{equation}
so that the statistics of the outcomes $M_A$ and $M_B$ selected by Diana's measurement results would violate (respect) the CHSH inequality (\ref{CHSHineq}). Here one has the matrices
\begin{eqnarray}
  \left({\rm Tr}_{AB}\left[Q^{}_A Q^{}_B\hat{\rho}_0\right]\right)_{xy,x'y'} &=& \frac{1}{4\sqrt{2}}
	\left(\begin{array}{cccc} 
	-1&0&0&1 \\
  0&1&1&0\\
  0&1&-1&0\\
  1&0&0&1 
  \end{array}\right), \nonumber\\ 
  \left({\rm Tr}_{AB}\left[P^{}_A Q^{}_B\hat{\rho}_0\right] \right)_{xy,x'y'} &=& \frac{1}{4\sqrt{2}}
	\left(\begin{array}{cccc} 
	-1&0&0&-1 \\
  0&-1&1&0\\
  0&1&1&0\\
  -1&0&0&1 
  \end{array}\right), \nonumber\\
  \left({\rm Tr}_{AB}\left[P^{}_A P^{}_B\hat{\rho}_0\right]\right)_{xy,x'y'}  &=& \frac{1}{4\sqrt{2}}
	\left(\begin{array}{cccc} 
	-1&0&0&1 \\
  0&-1&-1&0\\
  0&-1&1&0\\
  1&0&0&1 
  \end{array}\right), \nonumber\\ 
  \left({\rm Tr}_{AB}\left[Q^{}_A P^{}_B\hat{\rho}_0\right]\right)_{xy,x'y'} &=& \frac{1}{4\sqrt{2}}
	\left(\begin{array}{cccc} 
	1&0&0&1 \\
  0&-1&1&0\\
  0&1&1&0\\
  1&0&0&-1 
  \end{array}\right),
\end{eqnarray}
in the basis of $\{ |{\cal B}_{00}\rangle_{ab}, |{\cal B}_{01}\rangle_{ab}, |{\cal B}_{10}\rangle_{ab}, |{\cal B}_{11}\rangle_{ab}\}$,
with the choice of $Q^{}_A$, $P^{}_A$, $Q^{}_B$, and $P^{}_B$ above (\ref{CHSHBell}).

\subsection{Description by a moving observer}
\label{MovObsEntSwap}

For a bookkeeper with $t'$ slices represented by the green line in Figures \ref{NonLocalExpt} and \ref{EntSwapDLater} (right), Bob's measurement came first, then Diana's, and finally Alice's. This bookkeeper would describe the whole process as follows.

After Bob's measurement of the freely-chosen properties $Q_B$ or $P_B$, Diana is measuring the reduced state of photons $A$, $a$, and $b$, which is $\hat{\rho}'_1 = {\rm Tr}_{B}\, \hat{\rho}_0$. It is straightforward to see that, for all four Bell states of photons $a$ and $b$ obtained by Diana, the post-measurement state of photon $A$ is $\rho'_{2A} = (|0\rangle_A\langle 0| + |1\rangle_A \langle 1|)/2$, which is a mixed state at infinite temperature, indicating that photon $A$ is maximally entangled with photon $B$ or some qubit.
In contrast, if Diana obtains a separable states of photons $a$ and $b$, then the post-measurement state of photon $A$ will be a pure state depending on the outcome of Diana's measurement. 

In summary,
the descriptions in the time-order of $M_D$-$M_A$-$M_B$ and $M_D$-$M_B$-$M_A$ \footnote{In a real Bell test, most of the measurements $M_A$ and $M_B$ in a single run occur at different moments in the bookkeeper coordinates. This is why experimentalists have to introduce a time window to select coincidence detections \cite{AGR81, WJ98, GS15}.} in the conventional entanglement swapping 
(Section \ref{ExEntSwap}), $M_A$-$M_B$-$M_D$ and $M_B$-$M_A$-$M_D$ in the delayed-choice scenario (Section \ref{DelayedCh}), and $M_B$-$M_D$-$M_A$ and $M_A$-$M_D$-$M_B$ described by moving observers (Section \ref{MovObsEntSwap}), are all consistent with the measurement outcomes once the measurement events $M_A$, $M_B$, and $M_D$ are spacelike separated.

\section{Concluding remarks}

In this paper we have proposed and provided some basic analysis of quantum teleportation and entanglement swapping experiments between the ISS and LG. In our proposed experiments the measurement outcomes in a series by one agent can be causally independent of the measurement by the other agent in the same series.
The quantum information (QI) experiments with such a very long baseline also highlight the consistency of different time-ordered spacelike events described by different bookkeepers in their own reference frames.

Although quantum teleportation and entanglement swapping are much more difficult to achieve in laboratory than the Bell test because of more joint measurements and more participating objects, experimentalists have made great progress in the efficiency since the first implementation.
In 2002, the authors of Ref.\cite{JZ02} reported that the detected four-fold coincidence count rate is about 0.00065 per second. In 2012, the detected four-fold coincidence count rate is raised to 0.016 Hz \cite{MZ12}.
In 2015, the authors of Ref.\cite{HZ15} have achieved a four-fold rate of 100 Hz, though under that condition the Bell test is not feasible, and only the measurement of some entanglement witness can be done. Given the fact that the travel time of light from ISS to LG is about 1.3 second, if the four-fold rate can achieve 100 Hz or more (under high link losses with such a long baseline), 
the averaged fidelity of teleportation and the expectation values of the entanglement witness for entanglement swapping
could be convincing in these proposed experiment using ISS and LG.

Nevertheless, in these proposed low-efficiency, high-loss QI experiment in outer space, statistical bias could produce fake quantum advantage in the results (see Section \ref{Floophole}, corresponding to the loopholes in the Bell test \cite{Larsson14, DR05, DR07a, DR07b}).
It would be interesting if one could go beyond the averaged fidelity and find a new criterion, which is a counterpart of the CH inequality in the Bell test and is coincidence-time loophole-free \cite{CH74, LG14}, for quantum teleportation to address the quantum nonlocality issue there.

\ack
The authors thank Dr. Makan Mohageg for introducing them to the DSQL 
project and Prof. Thomas Jennewein for his very helpful comments and suggestions. They thank Prof. Paul Kwiat for his query on the loophole issue, and Prof. Jason Gallicchio for suggesting some standard references. SYL is supported by the Ministry of Science and Technology of Taiwan under Grant No. MOST 109-2112-M-018-002 and in part by the National Center for Theoretical Sciences, Taiwan. 
BLH is supported by NASA/JPL grant 301699-00001.

\appendix

\section{Bell test}
\label{Belltest}

In the Bell test, two agents Alice and Bob are located far apart in space. The third agent Charlie produces a pair of objects, sends one object to Alice, and the other to Bob. Each object carries two independent binary properties $Q$ and $P$, each property takes the value $+1$ or $-1$ after measurement. When receiving her object from Charlie, Alice freely chooses to locally measure its property $Q^{}_A$ or $P^{}_A$, and records the outcome of her local measurement. Roughly at the same coordinate time, Bob does a similar local measurement of $Q^{}_B$ or $P^{}_B$. After repeating this process for a few times, the bookkeeper may collect the record from both sides and test the CHSH inequality \cite{CHSH}
\begin{equation}
  \left| \, \langle Q^{}_A Q^{}_B\rangle +\langle P^{}_A Q^{}_B\rangle +\langle P^{}_A P^{}_B\rangle 
	-\langle Q^{}_A P^{}_B\rangle \, \right| \le 2 ,
\label{CHSHineq}
\end{equation}
which should always be respected in classical physics. Here $\langle .. \rangle$ represents an ensemble average of many single tests on the binary properties of $A$ and $B$.
Generalized to quantum physics, if the ensemble average of expectation values are taken with respect to a separable state of the object pair, then (\ref{CHSHineq}) still holds. However, entangled states of the objects can violate the CHSH inequality (\ref{CHSHineq}) by properly choosing the operators $Q^{}_A$, $P^{}_A$, $Q^{}_B$, and $P^{}_B$. For example, suppose a photon pair $A$ and $B$ is in the Bell state $|{\cal B}_{11}\rangle_{AB}$, then one may choose $Q^{}_A = \hat{\sigma}_z^{A}$ and $P^{}_A = \hat{\sigma}_x^{A}$ for $A$, $Q^{}_B = -\hat{\sigma}_+^{B} \equiv -(\hat{\sigma}_z^B+\hat{\sigma}_x^B)/\sqrt{2}$, and $P^{}_B =\hat{\sigma}_-^{B} \equiv  (\hat{\sigma}_z^B-\hat{\sigma}_x^B)/\sqrt{2}$ for $B$, so that
\begin{equation}
	\langle Q^{}_A Q^{}_B + P^{}_A Q^{}_B + P^{}_A P^{}_B - Q^{}_A P^{}_B \rangle^{}_{AB}	= 2\sqrt{2} >2, 
\label{CHSHBell}
\end{equation}
where $\langle O_1 O_2\rangle^{}_{AB} \equiv {}_{AB}\langle {\cal B}_{11} | O_1 O_2 |{\cal B}_{11}\rangle_{AB}$.

The choice of $Q^{}_A$, $P^{}_A$, $Q^{}_B$, and $P^{}_B$ depends on the entangled state which the expectation values are taken with respect to. For example, for $|{\cal B}_{00}\rangle_{AB}$, $|{\cal B}_{01}\rangle_{AB}$, or $|{\cal B}_{10}\rangle_{AB}$, one should choose 
$Q^{}_A = -\hat{\sigma}_z^A$ and/or $P^{}_A=-\hat{\sigma}_x^A$ instead to maximally violate the CHSH inequality. In the entanglement swapping experiment proposed in Section \ref{ExEntSwap}, the outcome of each of Diana's joint measurement is not predictable, and so the violation of the CHSH inequality can only emerge in data analysis {\it after} sufficiently many outcomes of all the relevant measurements are collected by the bookkeeper.

\section{``Before-before" and ``after-after" scenarios}
\label{BBAA}

Suppose Alice and Bob are moving apart from each other at the same speed, and Charlie at their mid-point emits a pair of photons, one photon goes to Alice, the other goes to Bob. Then, in Alice’s frame, her photon arrives before the other photon arrive at Bob, while in Bob’s frame, his photon arrives before the other photon meets Alice. In Ref. \cite{SS97}, the authors argued that Alice's device (detector or beam splitter) would accordingly act as receiving her photon before Bob does, while Bob's device would also act as receiving his photon before Alice does; this is called the ``before-before" scenario. Similarly, the case with Alice and Bob are moving towards each other is called the ``after-after" scenario in Ref. \cite{SS97}. In truth, these two scenarios are frivolous and nonexistent, because they are in contradiction with special relativity. 

The flaw in this way of thinking arises from not realizing that all physical observations are local. Alice can report on what she sees, not what Bob sees, not without some considerable amount of work, as follows: 
 
{\it a.} When Alice detects a photon, she does not know if Bob has ever detected anything or not. 
The best she can do at this moment is to determine the reduced state of her photon. 
 
{\it b.} Alice does not know if Bob detected a photon until she receives the message from Bob (or from a third agent) that Bob received his photon. 
In our setup with a very long baseline, this message can only arrive long after Alice’s detection. 
 
{\it c.} After Alice receives Bob's message, she will {\it infer} that Bob's detection occurred after her detection in terms of her coordinate time, which can be obtained by averaging her clock readings of 1) the departing time of a radar signal emitted from her to Bob's detection event, and 2) the arriving time of the echo of that signal from Bob's detection event, namely, in terms of the "radar time" of Bob's detection event \cite{DIn92, Lin20b}.
 
{\it d.} Only after Alice coordinatizes all the detection events in her past light cone could she do some quantum state tomography for the photon pair. This quantum state will be defined only in her frame, with a specific time-slicing scheme according to her radar time and radar distance, after a Hamiltonian has been constructed in the same frame to govern the continuous evolution of the photon pair between local measurements.
This could be called an ``after-before" (Alice-Bob) description, following Ref.\cite{SS97}. 
 
Same happens if we switch ``Alice" and ``Bob" in the above {\it a}.-{\it d}., resulting in a ``before-after" description in Bob's frame. Both the ``after-before" and ``before-after" descriptions are valid in each's own coordinates. The key observation is that quantum states depend on the choices of reference frame and gauge condition. 
They are not physical entities, just like the coordinate systems and gauge conditions. While in the above simple case the initial and final states in Alice and Bob's frames can look the same, only their measurement outcomes are physical. 
 
When one calls some phenomenon ``before-before" or ``after-after", one implies that both Alice and Bob’s frames are physical entities and they coexist for some unspecified yet physical observer. This simply cannot be done in special relativity, since physical consequences are independent of coordinate choices one uses for its description. 
All inertial observers are equivalent. All physical observers use the same means as Alice and Bob separately did to coordinatize a collection of events in spacetime in their reference frame.
In special relativity all observers are local, no super-observer can use {\it both} Alice and Bob’s frames at the same time.


\end{document}